\documentclass{article}
\usepackage[utf8]{inputenc}
\usepackage{amsmath,amsthm,amssymb}
\usepackage{tabularx}
\usepackage{bbding}
\usepackage{graphicx}
\graphicspath{ {./images/} }
\usepackage{indentfirst}
\usepackage{bm,multirow,float}
\usepackage{authblk}
\usepackage[a4paper, margin=1in]{geometry}
\usepackage{multirow}
\usepackage{url}

\title{Calculating the Expected Value of Sample Information in Practice: Considerations from Three Case Studies}
\author[1,2,3]{Anna Heath} 
\author[4,5,6,7]{Natalia R. Kunst}
\author[8]{Christopher Jackson}
\author[9]{Mark Strong}
\author[10]{Fernando Alarid-Escudero}
\author[11]{Jeremy D. Goldhaber-Fiebert}
\author[3]{Gianluca Baio}
\author[12]{Nicolas A. Menzies}
\author[13]{Hawre Jalal}
\author[ ]{on behalf of the Collaborative Network for Value of Information (ConVOI)}

\affil[1]{The Hospital for Sick Children}
\affil[2]{University of Toronto}
\affil[3]{University College London}
\affil[4]{University of Oslo}
\affil[5]{Yale University School of Medicine}
\affil[6]{Amsterdam UMC}
\affil[7]{LINK Medical Research}
\affil[8]{MRC Biostatistics Unit, University of Cambridge}
\affil[9]{School of Health and Related Research, University of Sheffield}
\affil[10]{Center for Research and Teaching in Economics (CIDE)}
\affil[11]{Stanford Health Policy, Centers for Health Policy and Primary Care and Outcomes Research, Stanford University}
\affil[12]{Harvard TH Chan School of Public Health}
\affil[13]{University of Pittsburgh}

\date{}
\begin{document}

\maketitle
\begin{abstract}
Investing efficiently in future research to improve policy decisions is an important goal. Expected Value of Sample Information (EVSI) can be used to select the specific design and sample size of a proposed study by assessing the benefit of a range of different studies.  Estimating EVSI with the standard nested Monte Carlo algorithm has a notoriously high computational burden, especially when using a complex decision model or when optimizing over study sample sizes and designs. Therefore, a number of more efficient EVSI approximation methods have been developed.  However, these approximation methods have not been compared and therefore their relative advantages and disadvantages are not clear.

A consortium of EVSI researchers, including the developers of several approximation methods, compared four EVSI methods using three previously published health economic models. The examples were chosen to represent a range of real-world contexts, including situations with multiple study outcomes, missing data, and data from an observational rather than a randomized study. The computational speed and accuracy of each method were compared, and the relative advantages and implementation challenges of the methods were highlighted.

In each example, the approximation methods took minutes or hours to achieve reasonably accurate EVSI estimates, whereas the traditional Monte Carlo method took weeks. Specific methods are particularly suited to problems where we wish to compare multiple proposed sample sizes, when the proposed sample size is large, or when the health economic model is computationally expensive. All the evaluated methods gave estimates similar to those given by traditional Monte Carlo, suggesting that EVSI can now be efficiently computed with confidence in realistic examples.
\end{abstract}

\section*{Introduction}

The Expected Value of Sample Information (EVSI) \cite{Schlaifer:1959,RaiffaSchlaifer:1961} quantifies the expected benefit of undertaking a potential future study that aims to reduce uncertainty about the parameters of a health economic model. The expected net benefit of sampling (ENBS), which is the difference between EVSI and the expected research study costs, can be used to inform decisions regarding study design and research prioritization. The future study with the highest ENBS should be prioritized if we wish to maximize economic efficiency. Thus, EVSI has the potential to determine the value of future research and to guide its design when accounting for economic constraints. 

Despite this potential, EVSI has rarely been used in practical settings for a variety of reasons \cite{Steutenetal:2013}. In the past, calculating EVSI in real-world scenarios has been based on nested Monte Carlo (MC) sampling \cite{Brennanetal:2007}, and this is computationally costly if we wish to produce accurate estimates with high precision. This computational burden is further increased when one aims to compute EVSI for multiple trial designs in order to determine the optimal (i.e., with the highest ENBS) research study \cite{ContiClaxton:2009,Jutkowitz:2019}. High performance computing resources can be used to overcome some of these barriers, but often at the expense of an increased requirement for programming skills and an increase in the complexity of the analysis. 

Several methods have been developed to overcome these computational barriers and unlock the potential of EVSI as a tool for research prioritization and trial design optimization \cite{Adesetal:2004,Weltonetal:2014,BrennanKharroubi:2007b,Strongetal:2015,Menzies:2016,Jalaletal:2015,JalalAlarid-Escudero:2017,BrennanKharroubi:2007,Heathetal:2017b,Heathetal:2018,Heathetal:2018b}. However, as many of these methods have been developed concurrently, they have not been compared. Additionally, EVSI estimation methods are typically evaluated using health economic models and trial designs chosen for computational convenience rather than those that reflect real-world decision making. 

Some of the EVSI estimation methods that have been proposed place restrictions on the structure of the underlying health economic model and/or the study design \cite{Adesetal:2004,Weltonetal:2014,BrennanKharroubi:2007b}. These restrictions typically take the form of an assumption about the study data that ensures that the prior and posterior model parameter distributions take the same form (conjugacy), and by doing so, allow for computationally efficient EVSI estimation. This, however restricts the applicability of these methods. EVSI estimation based on minimal modelling, where a comprehensive clinical trial is available to inform EVSI estimation, has also been proposed \cite{Meltzeretal:2011}. However, this paper aims to review EVSI estimation procedures for three case studies where the health economic models are based on a diverse evidence base and, when combined with the proposed study designs, do not fulfill the assumptions required for these restrictive or minimal modelling methods. 

Thus, our comparison is restricted to four recent calculation methods developed by (in chronological order) Strong et al.~\cite{Strongetal:2015}, Menzies \cite{Menzies:2016}, Jalal and Alarid-Escudero \cite{JalalAlarid-Escudero:2017} (extending a method proposed in Jalal et al.~\cite{Jalaletal:2015}), and Heath et al.~\cite{Heathetal:2017b,Heathetal:2018,Heathetal:2018b}. Whilst these methods are all based on different approaches and assumptions, they all provide estimation techniques for approximating EVSI that, in comparison to nested MC sampling methods, are less computationally demanding whilst retaining accuracy. 

Our primary goal is to test the four EVSI estimation methods across a range of health economic models and trial designs to gain a greater understanding of their behaviour in practice. We will evaluate the accuracy of EVSI estimation methods across the three models and the computational time required to obtain these estimates. These three models have several key features that reflect real-world trial design and may make it challenging to estimate EVSI in practice. These are: the presence of multiple trial outcomes, missingness or loss to follow-up in the data, and a study design that is observational rather than randomized.

\section*{Notation and Key Concepts}

Health economic decision making aims to determine the intervention, from some set of feasible alternatives, that is expected to be optimal in terms of utility (which is usually net monetary benefit or net health benefit  \cite{StinnettMullahy:1998}). We characterize a health economic model as a function that takes as an input a vector of parameters $\bm{\theta}$, and returns the costs and health effects associated with each intervention in the set of alternatives. Uncertainty in the input parameters is represented using a probability distribution $p(\bm\theta$). To find the optimal intervention, costs and effects are combined into a single measure of economic value by calculating the net benefit for each of the $T$ treatment options considered relevant, conditional on $\bm\theta$. Uncertainty about $\bm\theta$ induces uncertainty about the net benefit for each treatment $t=1,\dots,T$. We denote the net benefit for treatment $t$ given parameters $\bm\theta$ as $\mbox{NB}_t^{\bm\theta}$. Under the assumption of a rational, risk neutral decision maker, the optimal intervention given current evidence is the intervention associated with the maximum expected net benefit.

We consider that the model parameters can be split into two sets $\bm\theta=(\bm\phi,\bm\psi)$, where $\bm\phi$ is a subset of parameters that we wish to obtain more information on, and $\bm\psi$ are the remaining parameters. For example, clinical trials are informative for clinical outcomes but may not collect information about health state utilities or costs. The economic value of eliminating all uncertainty about $\bm\phi$ (assuming risk neutrality) is equal to the Expected Value of Partial Perfect Information (EVPPI) \cite{FelliHazen:1998,CoyleOakley:2008,Heathetal:2017}. This is given by
\begin{equation}
   \mbox{EVPPI} = \mbox{E}_{\bm\phi} \left[\max_{t}\mbox{E}_{\bm\theta\mid\bm\phi} \left[\mbox{NB}_{t}^{\bm\theta}\right]\right] - \max_{t} \mbox{E}_{\bm\theta}\left[\mbox{NB}_{t}^{\bm\theta}\right].\label{EVPPI}
\end{equation}

The EVSI is the value of collecting additional data, denoted $\bm X$, to inform the parameters $\bm\phi$ and is bounded above by the EVPPI. If these data had been collected and observed to have a value $\bm x$, they would be combined with the current evidence to generate an updated distribution for $\bm\phi$, $p(\bm\phi\mid\bm x)$. Under a Bayesian approach, this would in turn be used to update the distribution of the net benefit of each treatment. The optimal intervention conditional on the data $\bm x$ is the treatment associated with the maximum expected net benefit based on the updated knowledge about the relevant parameters $\bm\phi$. If the optimal intervention changes, compared to the current decision, then the information in $\bm x$ has value. However, as the data have not been collected yet (and may never be), the average value over all possible datasets is considered. Mathematically, EVSI is defined as
\begin{equation}
\mbox{EVSI} = \mbox{E}_{\bm X} \left[\max_{t} \mbox{E}_{\bm\theta\mid\bm X} \left[\mbox{NB}_{t}^{\bm \theta}\right]\right] - \max_{t} \mbox{E}_{\bm\theta} \left[\mbox{NB}_{t}^{\bm\theta}\right],\label{EVSI-eq}
\end{equation}
where the distribution of $\bm X$ can be defined through $p(\bm X, \bm\theta) = p(\bm\theta)p(\bm X\mid\bm\theta)$ where $p(\bm X\mid\bm\theta) = p(\bm X \mid\bm\phi)$ is the sampling distribution for the data given the parameters. We assume that the sampling distribution for the data is only defined conditional on $\bm\phi$, i.e., ~does not provide information on the value of the parameters $\bm\psi$, except through any relationship with $\bm\phi$.

\section*{Calculation Methods for EVSI}
It is rarely possible to compute EVSI analytically as the net benefit is often a complex function of $\bm\theta$. Additionally, it is challenging to compute the expectation of a maximum analytically as required in the first term of equation (\ref{EVSI-eq}). Therefore, a range of methods have been developed to approximate EVSI.

\subsection*{Nested Monte Carlo Computations for EVSI}
The simplest approximation method \cite{Brennanetal:2007} computes all the expectations in equation (\ref{EVSI-eq}) using MC simulation. The second term can be computed by simulating $s=1,\dots,S$ parameter values, $\bm\theta_s$, from $p(\bm\theta)$. The simulated values are used as inputs to a health economic model to obtain $S$ simulations of the net benefit for each intervention, denoted $\mbox{NB}_t^{\bm\theta_s}$. Note that this process is required to perform a ``probabilistic sensitivity analysis'' (PSA) \cite{BaioDawid:2011}, used to assess the impact of parametric uncertainty on the decision uncertainty, which is mandatory in various jurisdictions \cite{Eunethta:2014,Australia:2008,CADTH:2006}. The average of $\mbox{NB}_t^{\bm\theta_1},\dots, \mbox{NB}_t^{\bm\theta_S}$ for each intervention can be computed and $\max_{t} \mbox{E}_{\bm\theta} \left[\mbox{NB}_{t}^{\bm\theta}\right]$ is estimated by the maximum of these means. 

The first term in equation (\ref{EVSI-eq}) is more complex to compute by simulation. Firstly, $S$ datasets $\bm X_s$ must be generated conditional on the simulated $\bm\theta_s$ from the assumed sampling distribution $p(\bm X\mid\bm\theta_s)$. For each $\bm X_s$, we simulate $R$ values from the updated distribution of the model parameters $p(\bm\theta\mid\bm X_s)$. These $R$ simulations are used as inputs to the health economic model to simulate from the updated distribution of the net benefit for each intervention. The mean net benefit for each treatment option is then calculated to estimate $\mbox{E}_{\bm\theta\mid\bm X} \left[\mbox{NB}_{t}^{\bm\theta}\right]$
for $t = 1, \dots , T$. The maximum of these simulated means is then selected for each $\bm X_s$. Thus, to compute EVSI by MC simulation, we require $S\times R$ runs of the health economic model. This is computationally expensive for standard choices of $S$ and $R$, which are typically in the thousands. Therefore, the following methods focus on approximating the updated mean of the incremental net benefit associated with each intervention $t$ using a smaller simulation burden. We denote the expectation of the incremental net benefit, conditional on data $\bm X$, as \[\mu^{\bm X}_t = \mbox{E}_{\bm\theta\mid \bm X}\left[\mbox{NB}_t^{\bm \theta}\right].\] In a similar manner, we also denote the expectation of the incremental net benefit, conditional on some value of the parameters of interest $\phi$, as \[\mu^{\bm\phi}_t = \mbox{E}_{\bm\theta\mid\bm\phi}\left[\mbox{NB}_t^{\bm\theta}\right].\]

Finally, to increase the numerical stability of the following approximation methods, it is easier to work in terms of the incremental net benefit or loss, defined, without loss of generality, as $\mbox{INB}^{\bm\theta}_t = \mbox{NB}_t^{\bm\theta} - \mbox{NB}_1^{\bm\theta}$ for $t=2,\dots,T$. 

\subsection*{Strong et al.} 
The Strong \textit{et al.}~method estimates EVSI by fitting a regression model between the simulated values of the incremental net benefit, as the `dependent' or `response' variable, and a scalar or low-dimensional summary for the simulated dataset $\bm X$ as the `independent' or `predictor' variable(s) \cite{Strongetal:2015}. This low-dimensional summary for $\bm X$ should reflect how the data would be summarized if the study were to go ahead and must be computed for each simulated dataset $\bm X_s$. Once this regression model has been fitted, $\mu_t^{\bm X}$ is estimated by the fitted values from this regression model. EVSI is then estimated directly from these estimates of $\mu_t^{\bm X}$.

\subsection*{Menzies} 
Menzies \cite{Menzies:2016} presents two EVSI estimation methods, the most accurate of which estimates $\mu^{\bm X}_t$ by reweighting simulations of $\mu_{t}^{\bm\phi}$. This reweighting is based on the \emph{likelihood} of observing a simulated dataset $\bm X$ conditional on different values for $\bm\phi$. The term likelihood is used in the statistical sense and is equal to $p(\bm X\mid\bm\phi)$.

This method simulates $S$ future datasets $\bm X_s$ from $p(\bm X\mid\bm\phi_s)$. The likelihood for \emph{every} simulated vector for $\bm\phi$ is then calculated conditional on $\bm X_s$ . For the sample $\bm X_s$, $\mu^{\bm X_s}_t$ is estimated as the average of $\mu_{t}^{\phi}$, weighted by the likelihood of the dataset $\bm X_s$, and the method can therefore be seen as an example of importance sampling \cite{RobertCassella:2005,Rubin:1988}. EVSI is estimated based on the estimate of $\mu^{\bm X_s}_t$ for each future sample.

\subsection*{Jalal et al.} 
The Jalal \textit{et al.}~method published by Jalal and Alarid-Escudero \cite{JalalAlarid-Escudero:2017}, building on work from Jalal \textit{et al.}~ \cite{Jalaletal:2015}, fits a linear meta-model\footnote{A ``linear'' model is required for this method. However, non-linear functions of $\bm\phi$ can be defined and combined linearly to account for flexible relationships between the incremental net benefit and the parameters $\bm\phi$.} between the simulated incremental net benefit values, as the response variable, and simulations for $\bm\phi$, as the predictor variables. Each term of the linear meta-model is then rescaled based on a Gaussian-Gaussian Bayesian updating approach to estimate its ``posterior'' expectation across different future datasets $\bm X$. These estimated distributions are then recombined using the coefficients of the linear model to estimate $\mu^{\bm X}_t$ and compute EVSI.

For a proposed future data collection strategy of size $N$, the rescaling factor for each term of the linear meta-model is equal to \[\frac{N}{N+N_0},\] where $N_0$ is known as the prior effective sample size. In some prior-likelihood pairs, $N_0$ can be obtained analytically. In other settings, $N_0$ can be estimated using one of two estimation methods. If the data $\bm X$ can be summarized using a summary statistic $W(\bm{X})$, then $N_0$ can be computed as a function of the variance of $W(\bm X)$. If a suitable statistic cannot be derived, then nested posterior sampling can be used to estimate $N_0$. In this method, $S$ future datasets $\bm X_s$, $s=1,\dots, S$ are simulated. Each of these samples is used to update the information about the model parameters $p(\bm\theta\mid\bm X_s)$, typically using $R$ simulations and computing the mean for $\bm\phi$. The \emph{variance} of the mean for $\bm\phi$, across different samples $\bm X_s$, is then used to estimate $N_0$. Computationally, this nested sampling method to compute $N_0$ is relatively computationally expensive compared to the other two proposals to determine $N_0$. However, calculation of $N_0$ is only needed once to compute EVSI across study size.

\subsection*{Heath et al.} 
The Heath \textit{et al.}~\cite{Heathetal:2018,Heathetal:2018b} estimation method combines the simulations $\mu^{\bm\phi}_t$ and a modified nested MC sampling method to estimate EVSI. This method reduces the number of times the updated distribution of the net benefit must be simulated to estimate EVSI from $S$, typically at least 1000, to $Q$, usually between 30 and 50 \cite{Heathetal:2018b}. Thus, EVSI is estimated with $Q \times R$ health economic model runs.

The Heath \textit{et al.}~method uses nested MC sampling to estimate the \emph{variance} of the incremental net benefit for different future datasets. These estimated variances rescale simulations of $\mu^{\bm\phi}_t$ for $t=2,\dots,T$ to approximate simulations of $\mu^{\bm X}_t$ which can be used to estimate~EVSI. The Heath \textit{et al.}~method only requires a single nested simulation procedure to estimate EVSI across sample size \cite{Heathetal:2018b}.

\section*{Case Studies}
These EVSI methods are applied to three case studies designed to explore trial designs using health economic models that make EVSI estimation reflective of real-world decision making. The first case study is a stylized chemotherapy example used to evaluate EVSI estimation in the presence of multiple outcomes, reflecting a realistic trial design with a single primary, and multiple secondary, outcomes. The second case study evaluates EVSI methods in the presence of missingness in the data using a previously published health economic model to explore EVSI estimation when we account for standard considerations in trial design and development. Finally, we evaluate EVSI methods for a health economic model based on a time-dependent natural history model where the main data source is observational.  

\subsection*{Case Study 1: A New Chemotherapy Treatment}
This model was developed in Heath and Baio \cite{Heathetal:2018} to evaluate two chemotherapy interventions, i.e., the current standard of care and a novel treatment that reduces the number of adverse events. These two options are equal in their clinical outcomes so we focus on the adverse events. The probability of adverse events for the standard of care is denoted $\pi_{0}$ and $\rho$ denotes the proportional reduction in the probability of adverse events with the novel treatment.

All patients incur a treatment cost of \textsterling 110 for the standard of care or \textsterling 420 for the novel treatment. Patients without adverse events or those that have recovered have a quality of life (QoL) measure of $q$. The health economic impact of adverse events is modelled with a Markov model depicted in Figure 1. In this model, $\gamma_{1}$ and $\gamma_{2}$ denote the constant probability of requiring hospital care and dying, respectively, and $\lambda_{1}$ and $\lambda_{2}$ denote the constant probability of recovery given that an individual remains at home or enter hospital, respectively. The cycle length is 1 day and the time horizon is 15 days. Recovered patients incur no further cost while patients who die have a one-time cost of terminal care. There are costs and QoL measures associated with home and hospital care. PSA distributions for the model parameters are informed using previous data or defined using expert opinion with all distributional assumptions given in the supplementary material. 

\begin{figure}[ht]
\centering
\includegraphics[scale=0.8]{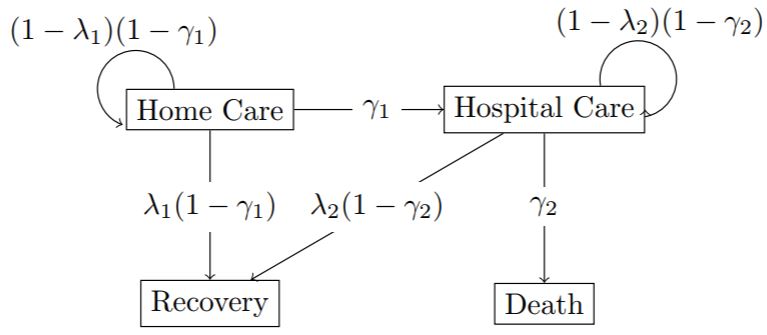}
\caption{A four state Markov model used to model the health economic impact of adverse events from a chemotherapy treatment.}
\label{fig:MarkovM}
\end{figure}

\subsubsection*{Sampling Distributions for $\bm{X}$} 
The EVSI is computed for a future two-arm randomized control trial whose primary outcome is the number of adverse events. As a secondary set of measures, the study monitors the treatment pathway for patients who experience adverse events. Thus, the trial directly informs six model parameters $\bm\phi = (\pi_{0}, \rho, \gamma_{1}, \gamma_{2}, \lambda_{1}, \lambda_{2})$ by collecting six outcomes. We will enrol 150 patients per arm.

To define the sampling distribution for the six outcomes, we model the number of adverse events using binomial distributions conditional on $\pi_{0}$ and $\rho$;
\[X_{AE_0} \sim Bin(150, \pi_{0})\,\, \mbox{and}\,\, X_{AE_1} \sim Bin(150, \rho \pi_{0}).\] The number of patients treated in hospital and the number of patients who die are modelled as \[X_{Hosp} \sim Bin(X_{AE_0} + X_{AE_1}, \gamma_{1})\,\, \mbox{and}\,\,X_{Death} \sim Bin(X_{Hosp}, \gamma_{2}).\] Finally, recovery time for patients who experience adverse events but recover is modelled with an exponential distribution conditional on the transition probabilities $\lambda_{1}$ and $\lambda_{2}$,
 \[T_{HC}^{i} \sim Exponential(\eta_{1})\] with $\eta_{1} = -\log(\lambda_{1}$) and $i = 1,\dots, X_{AE_0} +X_{AE_1} - X_{Hosp}$. The recovery time for every patient who recovers in hospital is modelled as \[T^j_{H}\sim Exponential(\eta_2)\] with $\eta_2 = -\log(\lambda_2)$ and $j=1,\dots,X_{Hosp}-X_{Death}$.

\subsection*{Case Study 2: A Model for Chronic Pain} 
This example uses a cost-effectiveness model developed by Sullivan et al. \cite{Sullivanetal:2016}, and extended in Heath \textit{et al.}~\cite{Heathetal:2018b}, to evaluate treatments for chronic pain. This is based on a Markov model with 10 states, where each state has an associated QoL and cost. The model is initiated when a cohort of patients receive their initial treatment for chronic pain. Patients can experience adverse events due to treatment and can withdraw from treatment due to adverse events or lack of efficacy. Following this, they can be offered an alternative therapy or withdraw from treatment. If they withdraw from this second line of treatment, they can receive further treatment or discontinue, both considered absorbing states as the model does not consider a death state.

As a treatment for chronic pain, a patient can first either be offered morphine or an innovative treatment. If they withdraw, they are offered oxycodone as an alternative treatment. Thus, the only difference between the two options is the first-line treatment where the innovative treatment is more effective, more expensive and causes fewer adverse events. A more in-depth presentation of all the model parameters is given in \cite{Sullivanetal:2016} where the parameter distributions are gamma for costs and beta for probabilities and utilities. The means of these distributions are informed by relevant studies identified following a literature review and the standard deviation is taken as 10\% of the underlying mean estimate. The per-person lifetime EVSI is calculated, assuming a discount factor of 0.03 per year over 15 years.

\subsubsection*{Sampling Distributions for  $\bm{X}$} 
EVSI is computed for a study that investigates the QoL weights for patients who remain on treatment without any adverse events and of patients who withdraw from the first line of treatment due to lack of efficacy. The individual level variability in these two QoL weights is modelled, for simplicity, as independent beta distributions although the assumption of independence may be invalid \cite{GoldhaberFiebertJalal:2016}. The population level mean QoL weight, i.e., ~the mean of the individual level QoL distribution, is defined as the value of those two health states in the Markov Model. The standard deviations of the individual level distributions is then set equal to 0.3, for patients who remain on treatment, and 0.31, for patients who withdraw due to lack of efficacy \cite{Ikenbergetal:2012}\footnote{This sampling distribution for the data causes some minor issues for the Gibbs sampling procedure used in the JAGS program for Bayesian updating.}. We compute EVSI for trials enrolling 10, 25, 50, 100 and 150 patients. We assume that only a proportion of the questionnaires are returned, leading to missingness in the data.

To generate the data, a response rate of 68.7\% is assumed, consistent with the  return rate observed in \cite{Gatesetal:2009}. We generate a response indicator for each patient in the trial using a Bernoulli distribution. If this indicator is 1, then we assume the patient returned the questionnaire and therefore we have observed utility scores for both states for that patient, simulated from the beta distributions specified above, conditional on the model parameters.

\subsection*{Case Study 3: A Model for Colorectal Cancer}
This example uses a health economic model developed by Alarid-Escudero \textit{et al.}~\cite{Alarid-Escuderoetal:2018} to evaluate a screening strategy for colorectal cancer (CRC) and pre-cancerous lesions known as adenomas. This model is based on a nine-state Markov model with age-dependent transition intensities which govern the onset of adenomas (pre-cancerous growths) and the risk of all-cause mortality. The onset of adenomas is modeled using a Weibull hazard conditional on age \[l(a) = \lambda_1g a^{g-1}\] where $\lambda_1$ and $g$ are the shape and scale parameters of the Weibull distribution and $a$ is the age of the patient. Model parameters are calibrated to observed literature and uncertainty in the model parameters $g$ and $\lambda_1$ reflects the uncertainty in these calibration targets.

The costs and QoL associated with each health state are used to evaluate the economic burden of CRC. The screening strategy is assumed to capture patients with adenomas and early cancer so they can be operated on before the cancer progresses and becomes clinically detected. The proposed screening strategy has a sensitivity with a mean of 0.98 and a specificity with a mean of 0.87. Some members of the general population have undiagnosed adenomas and early stage CRC at the onset of the simulation. 

\subsubsection*{Sampling Distributions for $\bm X$}
EVSI is computed for a study that investigates the onset of adenomas in the general population to inform the shape and scale of the Weibull hazard function. A cross-section of the general population aged between 25 and 90  without any screening history will be screened for the presence of adenomas with a gold standard test with 100\% sensitivity and specificity. Upon enrollment, the age of the subjects is recorded to determine the age-specific risk. EVSI is computed for trials enrolling 5, 40, 100, 200, 500, 750, 1000 and 1500 participants.

To generate prospective data, we simulate the enrolment age for participants. Demographic data from Canada in 2011, obtained from the Human Mortality Database \cite{HMD}, were used to generate study subjects with an age distribution representative of the general population, with ages restricted between 25 and 90 years. Conditional on their age $a$, a participant has a probability \[p(a) = 1 - e^{-\lambda_1a^g}\] of having an adenoma or CRC. The outcome for a specific subject was simulated from a Bernoulli distribution conditional on $p(a_i)$ \[X_i \sim Ber\left(p\left(a_i\right)\right)\] where $a_i$ is the age of participant $i$. We assumed that there is no missing data as participants are enrolled and undergo the test at the same clinic visit and no other data are collected. 

\subsection*{Analysis}
Comparing the presented EVSI estimation methods is challenging as their accuracy and computational time are dependent on choices made by the modeller and the computational efficiency of the method implementation. Table \ref{sim-choice} outlines the simulation choices that were made for the case studies. These choices were made to achieve EVSI estimates with a reasonable level of precision, while keeping the computation time manageable. For example, smaller sample sizes were necessary for models with a greater computational cost. We compared the speed and accuracy achievable by each method, and identified their relative advantages and challenges in practice.

 \begin{table}[!ht]
 \centering
 \begin{tabular}{|p{4.5cm}|p{2.5cm}|p{2.5cm}|p{2.5cm}|}
 \hline
 \multirow{2}{*}{Simulation Choices}& \multicolumn{3}{c|}{Case Study} \\
 \cline{2-4}
  & Chemotherapy side effects (1) & Chronic Pain (2) & CRC screening (3)\\
 \hline
  Initial PSA size & 100,000 & 100,000 & 5,000 \\   \hline
  Number of $\mu^{\bm\phi}_t$ simulations from EVPPI calculation & 100,000 & 100,000 & 5,000\\ \hline
  Nested simulation outer loop size & 100,000 &100,000 & NA\\ \hline
  Nested simulation inner loop size &100,000 &100,000 &NA\\ \hline
    Strong \textit{et al.}~sample size & 100,000 &100,000&5,000\\ 
    Menzies sample size & 20,000 &5,000 &2,500\\ \hline
      Jalal \textit{et al.}~$N_0$ computation method & nested posterior sampling &nested posterior sampling &nested posterior sampling \\ \hline
  Jalal \textit{et al.}~$N_0$ estimation outer loop size & 1,000 & 1,000 & 5,000\\ \hline
  Jalal \textit{et al.}~$N_0$ estimation inner loop size & 10,000 & 10,000 & 5,000 \\ \hline
  Jalal \textit{et al.}~$N_0$ estimation future sample size & 30 & 40 & 40 \\ \hline
  Heath \textit{et al.}~outer loop size & 50 & 50 & 50 \\ \hline
  Heath \textit{et al.}~inner loop size & 10,000 & 10,000 & 5,000 \\ \hline

 \hline
 \end{tabular}
  \caption{The simulation choices to compute EVSI for the four recent approximation methods and the nested MC method for case study 1, 2 and 3.}
 \label{sim-choice}
 \end{table}
The prior effective sample size for the Jalal \textit{et al.}~method needs to be computed once to estimate the EVSI across sample size. As posterior updating is slower for larger sample sizes, it is preferable to estimate $N_0$ with a small proposed sample $\bm X$. However, the estimation of $N_0$ also relies on a Gaussian approximation so the sample size of $\bm X$ should be sufficiently large to assume normality. Thus, the table above (Jalal \textit{et al.}~future sample size) highlights the sample size of $\bm X$ used in the nested posterior sampling to estimate $N_0$ that balances accuracy and computational speed.

For the first two case studies, we computed a standard error for the EVSI estimates by recomputing the EVSI 200 times, each time with the same PSA samples, so that this standard error reflects uncertainty arising from any simulation involved in the EVSI estimation procedure itself.

To obtain the computational time for the four recent approximation methods, computations were undertaken on a computer with an i7 Intel processor with 16 GB of RAM in \texttt{R} version 3.5.1. The nested MC computations were undertaken on a Linux Google Compute Engine virtual machine. The computation time give below is the total time across all cores. Code to undertake the computations in this paper is available from GitHub at \url{https://github.com/convoigroup/EVSI-in-practice}.

\section*{Results}
\subsection*{Case Study 1: Chemotherapy Side Effects} 

\begin{figure}[ht!]
\begin{center}
\includegraphics[width=\textwidth]{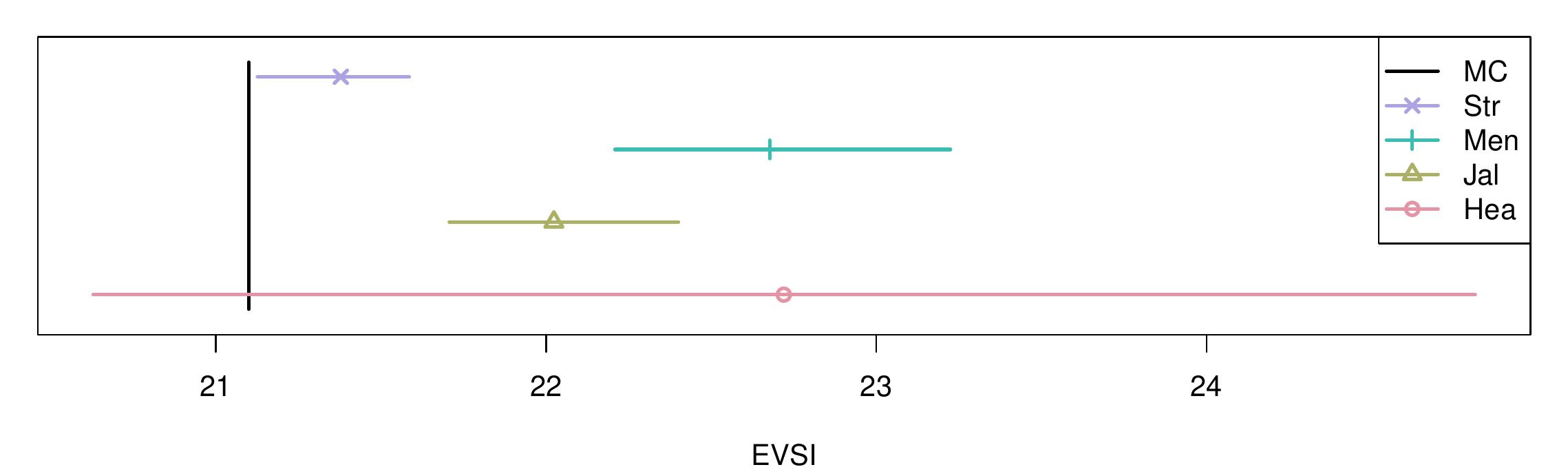}
\caption{The mean per-person EVSI estimates, across 200 simulated estimation procedures, for the five methods under consideration for the chemotherapy example with a future sample size of 150 and willingness-to-pay of $\pounds 30,000$. The 95\% central intervals from these 200 simulations are shown as horizontal lines and the gold standard MC estimator is shown as a vertical line.}
\label{Chemotherapy}
\end{center}
\end{figure}

Figure \ref{Chemotherapy} displays the 95\% central intervals for the four faster EVSI approximation methods, with the nested MC estimate shown as a vertical line. All the methods produce EVSI estimates that are relatively close to the EVSI estimated by nested MC sampling, which we assume is accurate given the large simulation size. The 95\% central interval for the Heath \textit{et al.}~method is the only interval that contains the ``true'' value, represented by the nested MC EVSI. At the same time, the Heath \textit{et al.}~estimate is associated with substantial variability compared to the other methods.

Implementing the Strong \textit{et al.}~and Jalal \textit{et al.}~methods involves finding a flexible regression model that fits well and is computationally feasible to estimate. As there are six parameters in this example, finding such a model was relatively challenging and required examination of residual plots.

\subsection*{Case Study 2: Chronic Pain} 
\begin{figure}[ht!]
\begin{center}
\includegraphics[width=\textwidth]{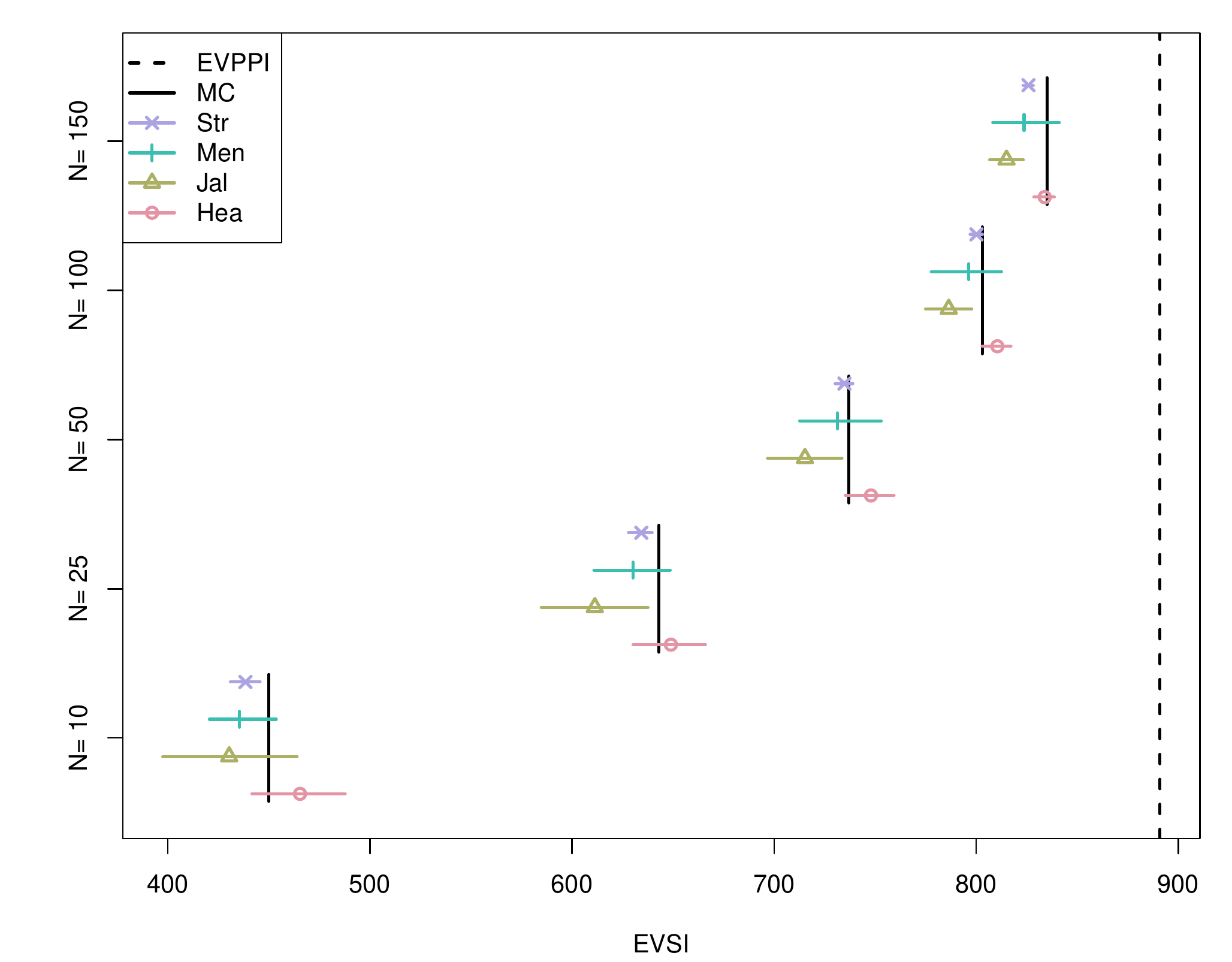}
\caption{The mean EVSI estimates, across 200 simulated estimation procedures, for the five methods under consideration for the chronic pain example. EVSI was calculated across  5 different sample sizes for the future trial. The 95\% central intervals from these 200 simulations are shown as horizontal lines and the gold standard MC estimator is shown as a vertical line.}
\label{chronic-pain}
\end{center}
\end{figure}

Figure \ref{chronic-pain} shows that the 95\% central intervals for the Heath \textit{et al.}~and the Menzies methods contain the nested MC estimate, which we assume to be accurate given the large simulation size, for all sample sizes. However, all methods are relatively close to the nested MC estimate. The Strong \textit{et al.}~ method produced the shortest 95\% central intervals while the three alternatives are relatively comparable. Note that the Menzies estimate is based on a smaller PSA simulation size but still offers similar variability compared to the other methods. 

In this example, the summary statistic used for the Strong \textit{et al.}~method is the geometric mean of $\bm{X}$ and $1-\bm{X}$. These statistics are sufficient to estimate the model parameters of the beta distribution and were derived using the Fisher-Neymann factorization theorem \cite{HoggCraig:1995}. Summarizing $\bm X$ using the arithmetic mean and variance gives incorrect EVSI estimates for this case study.

\subsection*{Case Study 3: Colorectal Cancer}
\begin{figure}[!h]
\includegraphics[width=\textwidth]{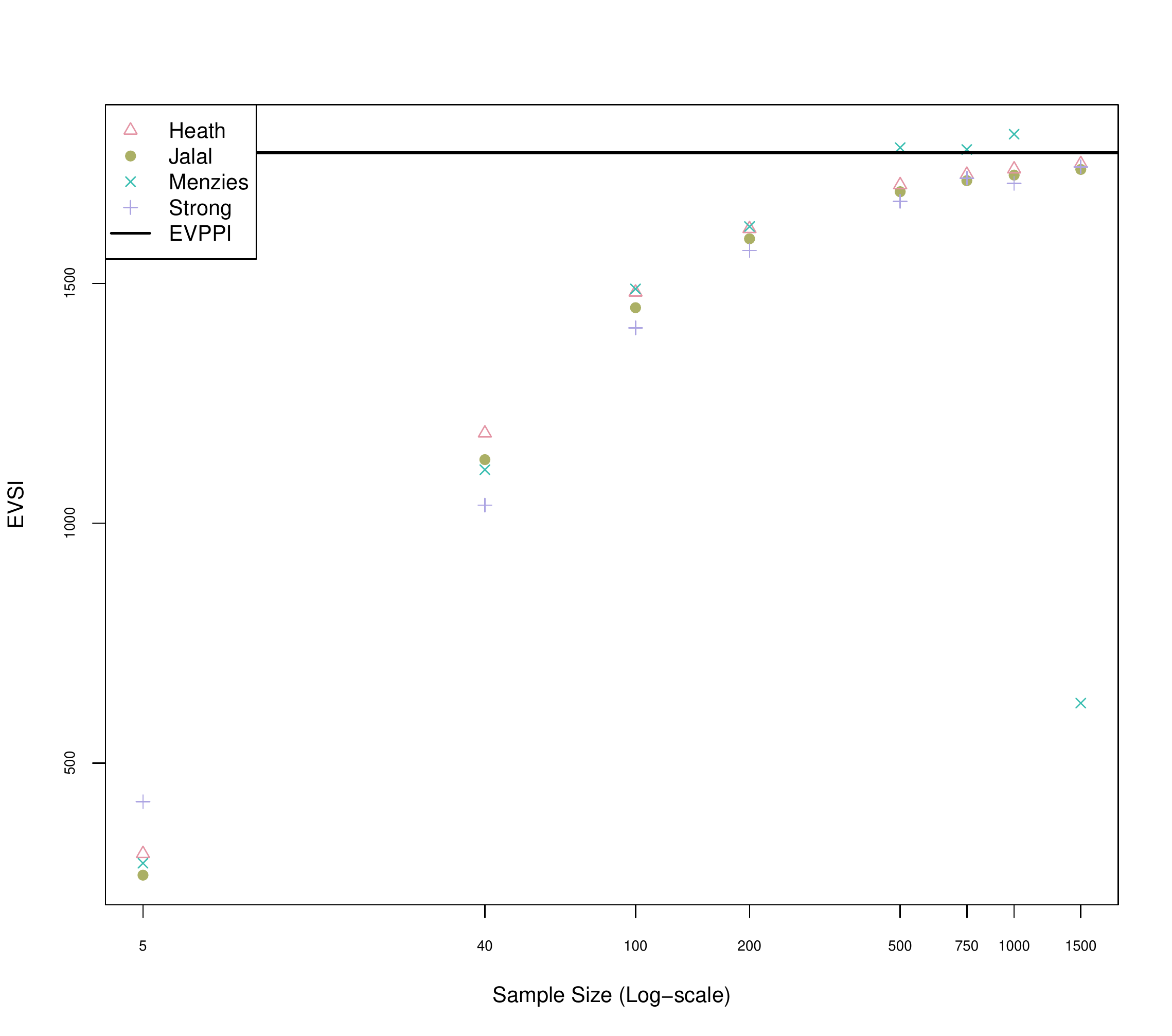}
\caption{EVSI estimates for the four methods under consideration for the CRC model. EVSI is calculated for 9 different sample sizes for the future trial and is plotted across sample size. The sample size is plotted on the log scale with the sample sizes marked on the natural scale. The EVPPI, computed using the Strong \textit{et al.}~EVPPI computation method \cite{StrongOakley:2014}, is included as a black line on this Figure.}
\label{Examp3Acc}
\end{figure}

Figure \ref{Examp3Acc} demonstrates a broad consensus among the four recent approximation methods for the CRC screening model. Nested MC simulations are not undertaken for this case study due to the computational time required to obtain suitably accurate estimates for comparison. Thus, while we can note that the four methods give similar results, we cannot assert that these EVSI estimates are ``correct.'' 

For a sample size of 1,500, the Menzies EVSI estimate is incorrect. This is because the likelihood tends to 0 for large sample sizes making the weighted samples difficult to approximate. Furthermore, the Menzies method slightly over-estimates the EVSI for sample sizes between 500 and 1000. This is because we only use a subset of the PSA simulations to obtain this EVSI estimate and the EVPPI, upper limit for EVSI, estimated using this subset is slightly over-estimated, judging from the full 5,000 PSA simulations.

\subsection*{Computational Time}
Table \ref{Computational-Time} shows the computational time for the five EVSI computation methods for each of the three case studies. For the first two case studies, all four alternatives are considerably faster than the nested MC method. For the third case study, the computational cost of the underlying CRC model meant that it was not computationally feasible to use the nested Monte Carlo method. 

For the first two case studies, the Heath \textit{et al.}~method has the lowest computation time as the underlying health economic model is fast. The Heath \textit{et al.}~method also estimates EVSI across multiple sample sizes simultaneously which improves the computational time for the Chronic Pain example compared to the Strong \textit{et al.}, and Menzies methods. For these two examples, the computation time required to fit an accurate regression model is relatively high, increasing the computation time for the Strong \textit{et al.}~method. The Jalal \textit{et al.}~method has the highest computation time as it uses nested MC simulation to calculate $N_{0}$. However, after estimating $N_{0}$, EVSI can be re-estimated for any sample size. Thus, if EVSI was to be estimated across more sample sizes, the Jalal \textit{et al.}~method would offer computational savings on the Strong \textit{et al.}, and Menzies methods. For the Chemotherapy example, the Menzies method has a similar computational cost to the other three methods. However, it is estimated based on a reduced simulation size; if all 100,000 PSA simulations are used, the computation time is greater than 2 hours. For the Chronic Pain example, the Menzies method is noticeably slower as the computation time for the likelihood increases when the proposed sample size of $\bm X$ is larger.
\begin{table}[ht!]
\begin{center}
\scalebox{0.85}{
\begin{tabular}{| p{3.5cm} | c | c | c | c | c |}
    \hline
    \multirow{2}{*}{Case Study} & \multicolumn{5}{c|}{Computational Time (mins)} \\ \cline{2-6}
     & Nested MC & Strong \textit{et al.} & Menzies & Jalal \textit{et al.} & Heath \textit{et al.}    \\
    \hline
    1: Chemotherapy & 60480 & 6.45 & 4.45 & 7.47 & 1.48 \\
    \hline
    2: Chronic Pain & 223200 & 12.05 & 86 & 22.27 & 2.46 \\
    \hline
    3: Colorectal Cancer & $*$ & 27.24 & 91 & 7.17 & 492  \\
    \hline
\end{tabular}}
\caption{The computational time required to produce EVSI estimates for the five methods under consideration for the three case studies presented in this review.}
\label{Computational-Time}
\end{center}
\end{table}

For the CRC screening example, the Jalal \textit{et al.}~method is fastest because, even though $N_0$ is estimated through nested MC simulation, it must only be computed once to estimate the EVSI across sample size. In contrast, for the Strong \textit{et al.}, method, $\bm X$ is summarized by finding the maximum likelihood estimates (MLE) for $g$ and $\lambda_1$ that must be estimated, using relatively slow computational optimization procedures, for each sample $\bm X_s$, $s=1,\dots,S$ and sample size. Thus, estimating the summary statistics is slow in this case study. The Heath \textit{et al.}~method is more computationally expensive as the underlying probabilistic sensitivity analysis for the CRC health economic model is expensive and must be rerun $Q\times S = 250,000$ times to compute EVSI. The computational time of the Menzies method is similar to the previous case studies.

\section*{Discussion}
The paper uses three case studies to assess four novel methods for approximating EVSI. These methods were developed in response to the immense computational burden required to estimate EVSI using nested MC simulations. As these methods were developed concurrently, no head-to-head comparison has been undertaken. Additionally, these methods have typically been assessed using health economic models designed for computational simplicity rather than reflecting real-life decision making.

Thus, we compared these four methods using case studies designed to cover a number of different trial designs, interventions and health economic model structures that may make the EVSI estimation more challenging. In general, the EVSI estimates were  accurate when the underlying assumptions for the respective methods were met, highlighting the importance of checking these assumptions. The computational complexity of these methods varies for different health economic models, different sampling distributions for the future data, and depending on whether optimization over different sample sizes is required. 

In general, we find that the four methods are comparable in terms of accuracy and computational time in these more realistic situations. However, it should be noted that appropriately assessing accuracy is challenging because differences in the EVSI estimate could lead to alternative future research recommendations, even when the difference is small. This is especially true for diseases with high incidence as the EVSI is multiplied by the incidence to determine whether the trial offers value for research investment. The determination of whether the EVSI is sufficiently precise will depend on the decision problem at hand, so care should be taken when interpreting these results. 

It is likely to be more useful to compare these methods on their ease of implementation. The ``optimal'' estimation method trading off accuracy, precision, computational time and ease of implementation will change depending on the health economic model structure, proposed trial design and analyst expertise. Due to the differences between these four methods and the inherent differences in health economic models and trial designs, giving general purpose recommendations is not simple and would not be unconditional. 

Nonetheless, this analysis highlights that the Strong \textit{et al.}~is accurate and efficient, provided the analyst can correctly summarize the trial data and fit a regression model. The Menzies method is accurate but computationally relatively expensive for large PSA simulation sizes. The Jalal \textit{et al.}~method is efficient when estimating EVSI across sample size but may require nested posterior sampling when considering realistic data collection exercises. Finally, the Heath \textit{et al.}~method is accurate and efficient when the health economic model has a low computation time but becomes more unfeasible as the model becomes more expensive. The Jalal \textit{et al.}~and Heath \textit{et al.}~methods required expertise in Bayesian methods for all the examples in this paper. 

While further research is required to give comprehensive guidance on the situations in which each of these methods is most useful, we can conclude that, provided the underlying assumptions of the method are met, any of the four methods chosen is likely to produce reasonable estimates in reasonable amount of time.

\section*{Contributions}
AH conceived the study, performed the analysis and drafted the paper; NRK advised on the study design and contributed to the analysis, results interpretation and drafting the paper; CJ advised on study design and contributed to drafting the paper; MS advised on the study design, verified the implementation of the Strong \textit{et al.}~method, and contributed to drafting the paper; FA-E advised on the study design, contributed to drafting the paper and verified the implementation of the Jalal \textit{et al.}~method; JDG-F advised on study design and contributed to results interpretation and drafting the paper; GB contributed to drafting the paper and the results interpretation; NM advised on the study design and contributed to drafting the paper; HJ conceived the study; contributed to drafting the paper and verified the implementation of the Jalal \textit{et al.}~method.  All authors approved the final draft.

\section*{Acknowledgements}
AH was funded by the Canadian Institute of Health Research through the SPOR Innovative Clinical Trial Multi-Year Grant . NRK was funded by the Research Council of Norway (276146) and LINK Medical Research. CJ was funded by the UK Medical Research Council programme MRC\_MC\_UU\_00002/11. This paper draws on work that MS conducted while supported by a NIHR Post-Doctoral Fellowship (PDF-2012-05-258) from 2013 to 2016. FA-E was funded by the National Cancer Institute (U01- CA-199335) as part of the Cancer Intervention and Surveillance Modeling Network (CISNET). JDG-F was funded in part by a grant from Stanford’s Precision Health and Integrated Diagnostics Center (PHIND). GB was partially funded by a research grant sponsored by Mapi/ICON at University College London. NM was supported by National Institutes of Health (NIH) [R01AI112438-02.]. HJ was funded by NIH/NCATS grant 1KL2TR0001856. The funding agreement ensured the authors' independence in designing the study, interpreting the data, writing, and publishing the report. The authors would also like to thank Alan Brennan, Michael Fairley, David Glynn, Howard Thom and Ed Wilson for their comments and discussion as part of the ConVOI group. 

\bibliographystyle{unsrt}
\bibliography{bib}

\appendix
\section{Inputs for the Chemotherapy Model}

\begin{table}[ht]
\begin{center}
\scalebox{0.65}{
\begin{tabular}{p{6cm}| p{2cm} | p{3cm} | p{3.1cm} | p{4.2cm}} 
     Model Input & Distribution & $1^{st}$ Prior Parameter & $2^{nd}$ Prior Parameter & Previous Data \\
     \hline
     $\pi_{0}$ - Probability of adverse events & Beta & 1 & 1 & Number of adverse events \\
     $\rho$ - Reduction in adverse events with treatment & Normal & Mean: 0.65 & Precision: 100 & No \\
     \textit{q} - QoL weight with no adverse events & Beta & 18.23 & 0.372 & No \\
     $\Gamma_{1}$ - Probability of hospitalization & Beta & 1 & 1 & Number of hospitalizations \\
     $\Gamma_{2}$ - Probability of death & Beta & 1 & 1 & Number of deaths \\
     $\gamma_{1}$ - Daily transition probability to hospital & $\frac{\Gamma_{1}}{15}$ & - & - & - \\
     $\gamma_{2}$  - Daily probability of death& $\frac{\Gamma_{2}}{15}$ & - & - & - \\
     $\lambda_{1}$ - Daily probability of recovery from home care& Beta & 5.12 & 6.26 & No \\
     $\lambda_{2}$  - Daily probability of recovery from hospital& Beta & 3.63 & 6.74 & No \\
    Cost of death& LogNormal & 8.33 & 0.13 & No \\
    Cost of home care & LogNormal & 7.74 & 0.039 & No \\
    Cost of hospitalization & LogNormal & 8.77 & 0.15 & No \\
    QoL weight for home care& Beta & 5.75 & 5.75 & No \\
    QoL weight for hospitalization & Beta & 0.87 & 3.47 & No \\
\end{tabular}}
\end{center}
\caption{The prior specification for the parameters underlying the Chemotherapy example including the distributional assumption and its parameters. Unless specified, the parameters are specified in the order used in the JAGS language for Bayesian updating. We indicate whether the stated prior is combined with data in the probabilistic health economic model.}
\label{Chemo-example-dist}
\end{table}
\end{document}